\documentclass[12pt,preprint]{aastex}

\shorttitle{IRAC Photometry of $\eta$ Cha}
\shortauthors{}

\bibpunct{(}{)}{,}{a}{}{;}
\bibpunct{(}{)}{,}{a}{}{;}

\begin{document}


\title{{\it Spitzer}/IRAC Photometry of the $\eta$ Chameleontis Association}


\author{S. T. Megeath\altaffilmark{1}, L. Hartmann\altaffilmark{2}, K. L. Luhman\altaffilmark{3}, \& G. G. Fazio\altaffilmark{1}}

\altaffiltext{1}{Harvard Smithsonian Center for Astrophyscis, MS-65,
  60 Garden St, Cambridge, MA 02138 (tmegeath@cfa.harvard.edu)}

\altaffiltext{2}{Dept. of Astronomy, U. Mich., 500 Church St., 830
  Dennison, Ann Arbor MI 48109}

\altaffiltext{3}{Department of Physics and Astronomy, Pennsylvania
  State University, University Park PA 16802 USA}



\begin{abstract}
We present IRAC 3.6, 4.5, 5.8 and 8~$\mu$m photometry for the
17 A, K and M type members of the $\eta$ Chameleontis
association.  These data show infrared excesses toward six of the
15 K and M stars, indicating the presence of circumstellar disks
around 40\% of the stars with masses of 0.1--1~M$_{\odot}$. The two
A-stars show no infrared excesses.  The excess emission around one of
the stars is comparable to the median excess for classical T~Tauri
stars in the Taurus association; the remaining five show comparatively
weak excess emission.  Taking into account published H$\alpha$
spectroscopy that shows that five of the six stars are accreting, we
argue that the disks with weak mid-infrared excesses are disks in
which the inner disks have been largely depleted of small grains by
grain growth, or, in one case, the small grains have settled to the
midplane.  This suggests that $\eta$~Cha has a much higher fraction
of disks caught in the act of transitioning into optically thin disks
than that measured in younger clusters and associations. 
\end{abstract}



\section{introduction}

The evolution of dusty circumstellar disks is currently the primary
observable constraint on planet formation around young stars.
Measurements of the properties of disks around stars in young clusters
and associations are especially useful since ages can be more reliably
determined for the clusters and associations than for individual
stars. By comparing the infrared emission from disks in
clusters/associations of different ages, disk evolution can be traced
as a function of age.  This method has been pioneered by ground-based
$JHKL$-band observations of young clusters, which can measure excess
infrared emission from small dust grains in the inner disk
\citep{lada00}.  By obtaining images at these wavelengths for rich
nearby clusters with ages ranging from $<$~1 to 30~Myr, \citet{hai01}
measured the fraction of stars that exhibited $L$-band excess emission
indicative of small dust grains in the inner regions of disks.  They
found that this fraction steadily decreased with cluster age with only
half of the stars showing $L$-band excesses by the age of 3~Myr. By
5~Myr, the $L$-band excesses are detected for only 12\% of the stars,
indicating that the small dust grains in most inner disks have been
depleted. The one significant exception may be the 5--9~Myr
$\eta$~Chameleontis association \citep[hereafter
  ECA;][]{mam99,law01,ls04}; \citet{lyo03} reported $L$-band excesses
in 60\% of the stars in this association.  However, \citet{hai05}
detected $L\arcmin$-band excesses for only 17\% stars in the ECA,
which is consistent with the $L$-band excess fractions measured for
other clusters by \citet{hai01}.

To resolve this inconsistency and obtain a more robust estimate of the
fraction of stars with disks in the ECA, we have obtained photometry
at wavelengths longer than the previous ground-based $JHKL$
measurements by using the Infrared Array Camera \citep[IRAC;][]{faz04}
on the {\it Spitzer Space Telescope}. This Letter describes these
observations, identifies the ECA members that have disks based on the
IRAC photometry, and compares the resulting disk frequency to previous
measurements of this association and other young associations and
clusters.

\section{Observations}

As a part of the Guaranteed Time Observation program ``Disk Evolution
in the Planet Formation Epoch'', we obtained images at 3.6, 4.5, 5.8,
and 8.0~\micron\ of the 15 known K- and M-type and two known A-type
members of the ECA \citep{mam99,law02,lyo04,sz04,ls04}.  IRAC observed
the targets during a 1 hour period ranging from Julian date
2453167.068 to 2453167.111 and obtained 0.6 and 12 second exposures at
three dithered position toward each of the targets.  We extracted the
photometry from the Basic Calibrated Data products (pipeline version
S11.4.0) using the IDLPHOT package (Landsman 1993).  The IDLPHOT
routines were integrated into a custom IDL program that uses the world
coordinate system information in the image headers to identify the
images containing a given star, find the star within the images, and
extract the photometry.  A 5 pixel radius aperture was used for each
star, with a sky annulus extending from 5 to 10 pixels.  We used zero
points (for units of DN s$^{-1}$) of 19.6642, 18.9276, 16.8468 and
17.3909 \citep{reach05} and aperture corrections of 1.061, 1.064,
1.067, and 1.089 for the [3.6], [4.5], [5.8] and [8.0] bands,
respectively.  The photometry was then corrected using position
dependent gain maps available on the {\it Spitzer} Science Center Web
pages.  The results are in summarized in Table~1.

\section{Results}

Mid-infrared colors constructed from the IRAC photometry are a
sensitive probe of infrared emission from dusty disks around young
stars.  Using both models and observations of well-studied classical
T-Tauri stars, \citet{allen04} and \citet{hartmann05} have studied the
color of young stars in the IRAC color-color diagram ($[3.6]-[4.5]$
vs.  $[5.8]-[8.0]$) and have found that young stars with disks have
distinctive mid-infrared colors.  Fig.~1 shows the IRAC color-color
diagram for the15 stars in the ECA.  Six stars show significant
infrared excesses in their $[5.8]-[8.0]$ colors, indicating that 6/15
stars with masses of 1--0.1~M$_{\odot}$ have disks. The two A stars,
HD 75505 and the double RS Cha, do not show an excess ($[5.8]-[8.0]
\sim 0.0$ for both stars).  Infrared excess emission was recently
detected in {\it Spitzer} 24~$\mu$m photometry of the one remaining
member which we did not image in our IRAC survey: the B8 star $\eta$
Cha.  This star appears to have a debris disk \citep{rieke2005}.  In
total, 7/18 of the member stars in the ECA currently show evidence for
disks, although longer wavelength observations may increase this
ratio.

The initial estimate of the excess frequency was made by
\citet{lyo03}, who obtained $JHK$-band and $L$-band (3.5~$\mu$m)
photometry of 11 of the K and M-type stars and concluded that 7/11 of
those stars had infrared excesses on the basis of their $K-L$ colors.
More recently, \citet{hai05} combined $JHK_s$ 2MASS photometry with
VLT $L'$-band (3.8~$\mu$m) photometry.  By using the spectral types of
\citet{ls04} to determine the intrinsic $K_s-L\arcmin$ colors, they
found that only 2/12 of the stars had significant infrared excesses.
Although we find that 6/15 of these stars have significant infrared
excesses in the IRAC bands, the $K_s-[3.6]$ color, which is similar to
the $K_s-L$ color, only shows significant excesses for two of the
sources: stars 11 and 15 \citep[Fig.~2;][]{ls04}. These are the two
infrared excess sources identified by \citet{hai05}.  At longer
wavelengths, stars 5 and 9 show infrared excesses, consistent with the
results of \citet{lyo03}; however, these sources do not show
substantial excesses in the $K_s-[3.6]$ color.  Finally, sources 3, 6,
and 12, which were reported to have excesses by \citet{lyo03}, do not
show infrared excesses in any of the IRAC bands.  The IRAC results
agree with the conclusion of \citet{hai05}: offsets and scatter in
their $K$ and $L$--band photometry led \citet{lyo03} to overestimate
the number of stars with $L$--band excesses.  We obtain a higher
fraction of stars with excesses than \citet{hai05}, but only because
most of the stars with disks show significant IR excesses only at
wavelengths longward of 4~$\mu$m.

\begin{table}
\begin{center}
\caption{IRAC Photometry \label{table}}
\begin{tabular}{llrrrrl}
\tableline\tableline
ID$^1$ & Spect.& 3.6~$\mu$m & 4.5~$\mu$m & 5.8~$\mu$m & 8.0~$\mu$m & Other\\
  & Type$^2$ & mag. (unc)$^3$ & mag. (unc) & mag. (unc) & mag. (unc) & Designation \\
  1 &    K6 &    7.17 ( 0.02)  &    7.20 ( 0.02)  &    7.16 ( 0.01)  &    7.11 ( 0.02) & RECX 1\\        
  3 & M3.25 &    9.27 ( 0.02)  &    9.21 ( 0.02)  &    9.09 ( 0.05)  &    9.15 ( 0.01) & RECX 3\\        
  4 & M1.75 &    8.45 ( 0.02)  &    8.45 ( 0.02)  &    8.37 ( 0.01)  &    8.32 ( 0.02) & RECX 4\\        
  5 &    M4 &    9.59 ( 0.03)  &    9.50 ( 0.03)  &    9.37 ( 0.01)  &    8.89 ( 0.01) & RECX 5\\        
  6 &    M3 &    9.15 ( 0.04)  &    9.07 ( 0.07)  &    9.04 ( 0.01)  &    9.04 ( 0.01) & RECX 6\\        
  7 &    K6 &    7.52 ( 0.01)  &    7.54 ( 0.02)  &    7.49 ( 0.01)  &    7.46 ( 0.01) & RECX 7\\        
  8 &    A7 &       - (   - )  &      -  (   - )  &    5.40 ( 0.01)  &    5.40 ( 0.01) & RS Cha AB\\     
  9 &  M4.5 &    8.99 ( 0.02)  &    8.80 ( 0.01)  &    8.57 ( 0.01)  &    7.97 ( 0.02) & RECX 9\\        
 10 &    M1 &    8.58 ( 0.01)  &    8.61 ( 0.01)  &    8.55 ( 0.02)  &    8.49 ( 0.01) & RECX 7\\        
 11 &  K5.5 &    7.09 ( 0.02)  &    6.86 ( 0.01)  &    6.57 ( 0.01)  &    5.97 ( 0.01) & RECX 11\\       
 12 & M3.25 &    8.19 ( 0.01)  &    8.15 ( 0.01)  &    8.10 ( 0.02)  &    8.07 ( 0.01) & RECX 12\\       
 13 &    A1 &    6.97 ( 0.01)  &    6.97 ( 0.01)  &    6.97 ( 0.01)  &    6.97 ( 0.01) & HD~75505\\    
 14 & M4.75 &   10.53 ( 0.01)  &   10.26 ( 0.05)  &   10.04 ( 0.01)  &    9.48 ( 0.01) & J0841.5-7853 \\  
 15 & M3.25 &    8.38 ( 0.03)  &    7.91 ( 0.01)  &    7.42 ( 0.03)  &    6.51 ( 0.01) & J0843.3-7905\\  
 16 & M5.75 &   11.02 ( 0.01)  &   10.75 ( 0.01)  &   10.42 ( 0.01)  &    9.76 ( 0.01) & J0844.2-7833\\  
 17 & M5.25 &   10.07 ( 0.02)  &   10.00 ( 0.01)  &    9.92 ( 0.01)  &    9.90 ( 0.01) & J0838.9-7916\\  
 18 &  M5.5 &   10.57 ( 0.03)  &   10.45 ( 0.01)  &   10.40 ( 0.01)  &   10.37 ( 0.01)  & J0836.2-7908 \\

\tableline
\tablenotetext{1} {Identification numbers from
\citet{ls04}.}
\tablenotetext{2}{Spectral types from \citep{ls04}.}
\tablenotetext{3}{The uncertainties are the standard deviation
of the magnitudes obtained for each set of three frames.}
\end{tabular}
\end{center}
\end{table}

\section{Discussion}

The ECA provides an important opportunity to study disk evolution in
an age range (5-9 Myr) that has been poorly probed so far.  Our
finding that 40\% of the systems with masses of 0.1--1~M$_\odot$ show
dust emission from disks is very similar to the frequency found using
{\it Spitzer} photometry of stars with similar masses in the $\sim 4$~
Myr-old cluster Tr 37 \citep[e.g.,][]{aguilar2005}, but much higher
than the $\sim 10$\% frequency found using $N$-band photometry of the
10 Myr-old TW Hya association \citep{rayjay1999} or the $< 10$\%
frequency found using {\it Spitzer} photometry for the $\sim
12$~Myr-old cluster NGC 7160 \citep{aguilar2005}.  The two stars with
detectable excesses At 3.6~$\mu$m yields a frequency similar to the
$\sim 12$\% $L$-excess frequency found for the $\sim 5$~Myr-old
cluster NGC 2362 \citep{hai01}.  These observations illustrate the
importance of longer wavelength observations in detecting disks and
the necessity of comparing infrared excess frequencies only within
similar wavelength bands.  These data also suggest that the disks in
small associations like the ECA evolve similarly to disks in large
associations with massive stars, such as Tr~37.

Of the stars with IR excesses, one (star 15) has an infrared spectral
energy distribution that is essentially the same as that of a typical
classical T Tauri star (CTTS) with an optically thick disk extending
inward close to the stellar surface (Fig.~3).  The other four show
greatly reduced or absent infrared excess emission, particularly at
wavelengths $\lesssim 6 \mu$m.  This is apparent in both the
$[3.6]-[4.5]$ and $[5.8]-[8]$ colors as compared to the colors of CTTS
in Taurus (Fig.~1) and the spectral energy distributions
compared to the median Taurus spectral energy distribution (Fig.~3).

The ECA is remarkable for the relatively large fraction of stars
exhibiting both disk gas accretion, although at a reduced rate, and
very reduced hot dust emission (Fig~4).  \citet{law04} inferred that
stars 5, 9, 11, and 15 are accreting on the basis of H$\alpha$
profiles.  The H$\alpha$ emission--line equivalent widths of stars 14
and 16 are $\-12$ and $-58$\AA, respectively \citep{law02,sz04};
the analysis of \citet{white2003} suggests that star 16 is accreting
while star 14 probably just exhibits chromospheric H$\alpha$ emission.
Thus, 5 of the 6 stars with IRAC excesses exhibit gaseous accretion
detectable in H$\alpha$.

While longer-wavelength observations will be needed to assess outer
disk properties, the presence of gas accretion leads us to suspect
that the four ECA stars with weak short-wavelength excesses
(5, 9, 14, and 16) are ``transition'' T Tauri disk systems such as TW Hya
\citep{calvet2002,uchida2004} and CoKu Tau 4
\citep{forrest2004,dalessio2005}. These disks are thought to be
transitioning from optically thick accretion disks into optically thin
planetesimal and debris disks; in this transition phase the inner
disks are optically thin and the outer disk is still optically thick.
Transition disks exhibit very low or no dust emission at shorter
wavelengths and strong emission at wavelengths $\gtrsim 10 \mu$m,
implying the presence of an inner disk ``hole'' largely cleared of
small dust grains and a substantial outer disk of gas and dust.  In
such disks, the dust detected by IRAC is transported into the inner
disk by accretion from the outer disk.  In contrast, ``debris disk''
systems, in which small dust grains are produced by the collisions of
planetesimals, rarely show excesses at wavelengths $\lesssim 10 \mu$m,
and they do not show evidence of significant continuous gas accretion onto
the central star \citep[as distinct from infalling bodies; e.g.][and
  references therein]{grady2000}.  Star 11, which does show a
significant 3.6~$\mu$m excess, may have an optically thick flat disk
due to dust settling and or grain growth in the inner disk \citep{calvet2005a}.

Current {\it Spitzer} Infrared Spectrometer (IRS) studies indicate
that the number of transition disk systems in Taurus (ages $\sim
2$~Myr) is relatively small, $\lesssim 5\%$ of the number of classical
T-Tauri star/disk systems \citep[][]{calvet2005b}.  If all CTTS in
Taurus pass through a transition disk stage, this suggests that the
transition phase lasts 5\% of the total disk lifetime.  In sharp
contrast, the ECA appears to show a transition disk fraction of 4/6 of
all systems with infrared excesses.  Although the ECA contains a
higher fraction of low mass mid-M type stars than Taurus, of the eight
CTTS and 10 weak-line T-Tauri stars with spectral types of M3--6 in
Taurus, none show transition disks.  This suggests that the difference
between the ECA and Taurus is not due to the different distribution of
masses in the two assocations.  Whether this remarkable difference
results from a slowing of disk transitions with increasing age, or
some other, possibly environmental, factor is not clear.

We now estimate the size and gas surface density in the inner disk
holes.  To estimate the sizes of the inner disk holes, we scale a
model for TW Hya, which has a significant dust excess in the IRAC [8]
band but little at shorter wavelengths, and has been modelled as
having a disk ``wall'' at 4 AU with a relatively evacuated region at
smaller radii \citep{calvet2002}.  TW Hya has a luminosity of about
$0.25 L_{\odot}$ \citep{webb1999} and the ECA disk systems have
luminosities of $\sim 0.1--0.02 L_{\odot}$, so our observations
suggest evacuated regions of $\sim 2--0.4$ AU in size.

The presence of accretion indicates that the inner disk hole does
contain gas, and the amount of gas in the inner disk hole can be
estimated from the accretion rates inferred from the H$\alpha$
velocity widths (Fig.~4).  \citet{law04} estimated mass accretion
rates of $\sim 10^{-10.4} M_{\odot}~yr^{-1}$ for stars 5, 9, and 11,
and an accretion rate of $\sim 10^{-9} M_{\odot}~yr^{-1}$ for 15.  The
lowest rates are well below those observed in Taurus and other young
regions \citep[see summary in][]{calvet2005a}. This is not surprising,
given the advanced age of ECA and the observational indications of
decreasing accretion rates with age, although the estimated ECA
accretion rates appear to fall a bit below a simple model of a
viscously evolving disk \citep{calvet2005a}.

Extrapolation of the steady accretion disk models by
\citet{dalessio1999} to somewhat lower stellar luminosities and masses
suggests that at accretion rates of $\sim
10^{-10}$~M$_{\odot}$~yr$^{-1}$, the disk gas surface density would be
of order 1~g~cm$^{-2}$ at 1 AU and scale roughly as $R^{-1}$ for a
typical viscosity parameter of $\alpha = 10^{-2}$.  Taking a dust
opacity (per gram of gas) for interstellar (ISM) dust in the IRAC
4.5~$\mu$m band of order $9 \, {\rm cm^2 g^{-1}}$ \citep[derived
  from][]{indeb2005}, a vertical optical depth of unity requires a
surface density of $\sim 0.1~{\rm g~cm^{-2}}$. For ISM dust, the
accretion rates imply a 4.5~$\mu$m optical depth of order 10 at 1 AU
and about 100 at 0.1 AU.  To make the disk optically thin within the
0.1 - 1 AU range, as required by the observations of the ECA
transition disks, the opacity would have to be reduced by $\sim 10^1 -
10^2$ from the ISM case.  While there are substantial uncertainties in
several parameters, this result suggests that the dust opacity per
unit gas mass in the ECA transition disks has declined substantially
from the initial value.  Thus, it appears likely that the ECA has an
unusually large proportion of stars in which the inner disks have had
substantial dust growth, reducing mid-infrared opacities to low values
while still maintaining a significant flow of accreting gas at radii
$\lesssim 1$~AU.  These data are further evidence that in as least
some disks, the primodial dust grains in the inner disk are collected
into larger dust grains and planetestimals in the first 10~Myr of
stellar evolution, while gas is still is present in the disk.

\acknowledgments

 This work is based on observations made with the {\it Spitzer} Space
 Telescope, which is operated by the Jet Propulsion Laboratory,
 California Institute of Technology, under NASA contract 1407.
 Support for this work was provided by NASA through contract Number
 1256790 issued by JPL/Caltech.  K. L. was supported by grant
 NAG5-11627 from the NASA Long-Term Space Astrophysics program.
 L. H. was supported by the NASA grant NAG5-13210.  2MASS is a joint
 project of the University of Massachussetts and Infrared Processing
 and Analysis Center/California Institute of Technology, funded by
 NASA and the National Science Foundation. We thank the referee
Warrick Lawson for his valuable comments.

{}

\begin{figure}
\epsscale{1.}
\plotone{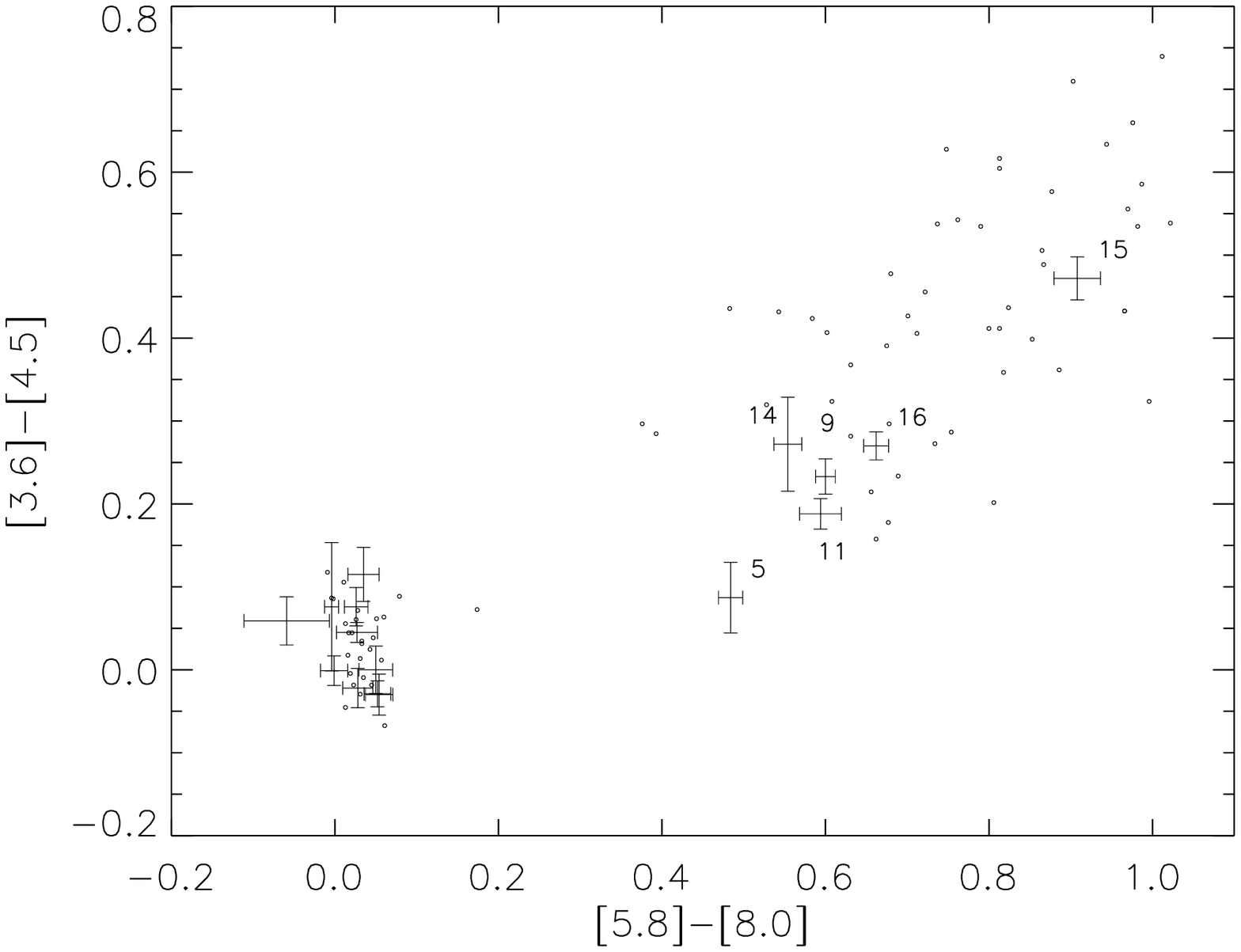}
\caption{IRAC color-color diagram for the 15 known K and M type
  members of the $\eta$ Cha association and the A-star HD~75505.  The
  dots shows the colors of pre-main-sequence stars observed in the
  Taurus regions \citep{hartmann05}.  Pure photospheres are clustered
  around (0,0), with the M stars showing colors of $[5.8]-[8.0]
  \approx 0.03$ and $[3.6]-[4.5]$ ranging from $0.15$ to $-0.05$.  The
  excess sources are labeled using the identification numbers from
  \citet{ls04}.}
\end{figure}

\begin{figure}
\epsscale{1.}
\plotone{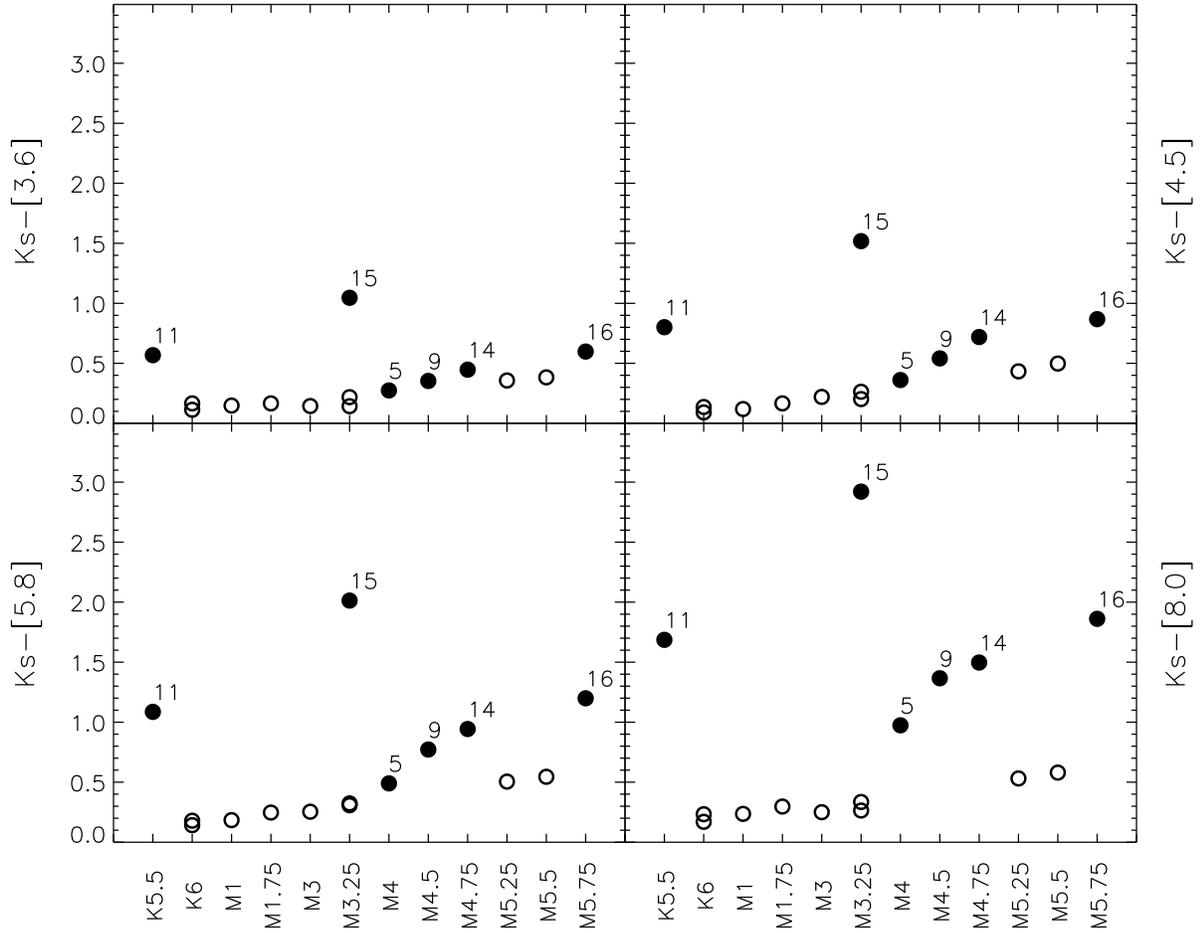}
\caption{Infrared excess as a function of spectral type using the
  spectral types from \citet{ls04}.  We display the color for all four
  IRAC bands.  The filled circles are infared excess sources, the open
  circles do not show excesses. The $K_s$ magnitudes are taken from the
  2MASS point source catalog.}
\end{figure}

\begin{figure}
\epsscale{1.}
\plotone{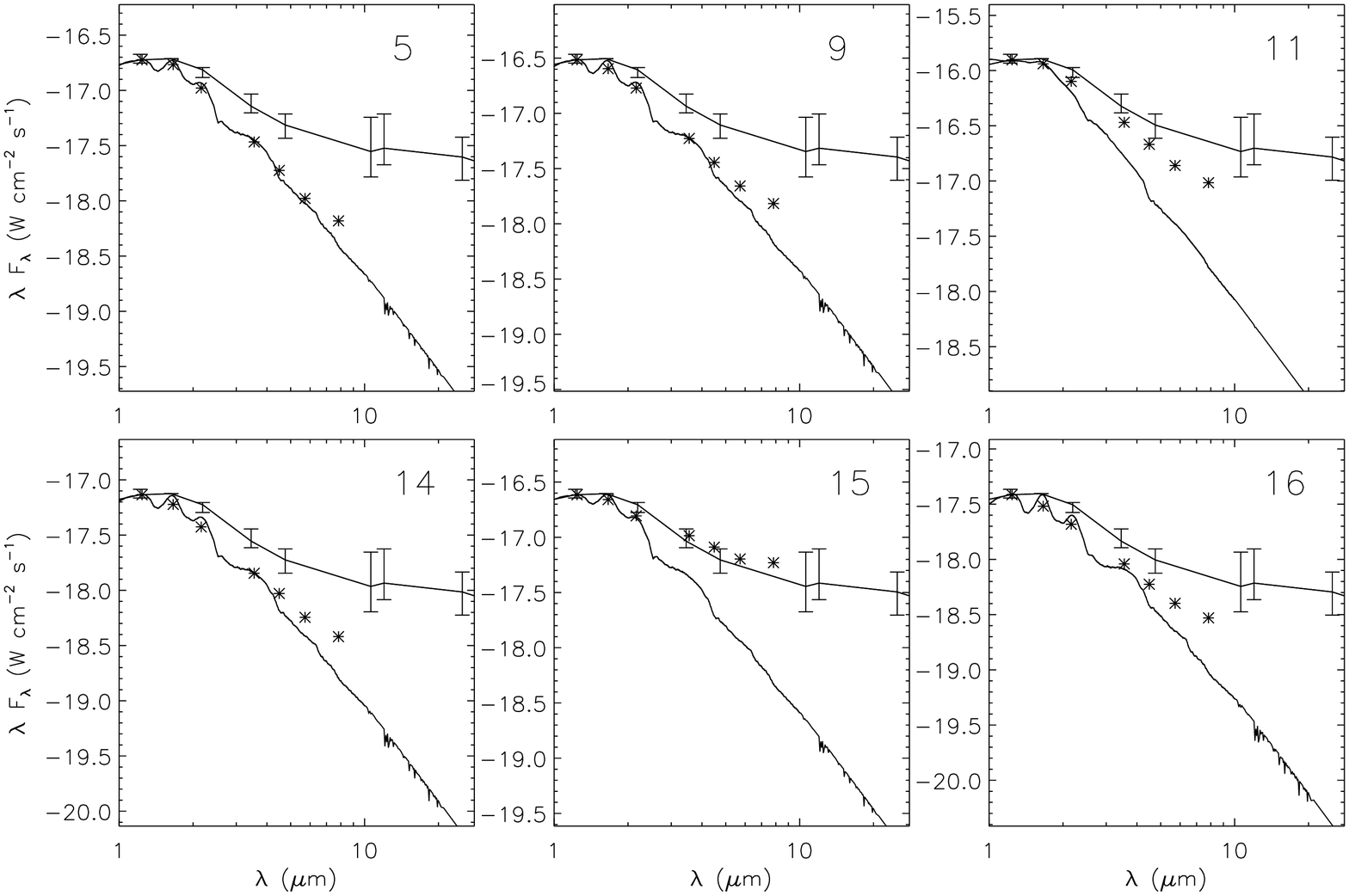}
\caption{Spectral energy distributions for the six infrared excess
  sources.  The $J$, $H$ and $K_s$ fluxes are from the 2MASS point
  source catalog. The upper line shows the median classical T-Tauri
  locus from \citet{dalessio1999}.  For comparison, we show the NextGen
  models for stars with effective temperatures from \citet{ls04} and a surface
  gravity of $g = 4.0$ \citep{haus1999}; these models are smoothed to
  a resolution of 0.1~$\mu$m and are scaled to the $J$-band flux of
  the observed stars.}.
\end{figure}

\begin{figure}
\epsscale{1.}
\plotone{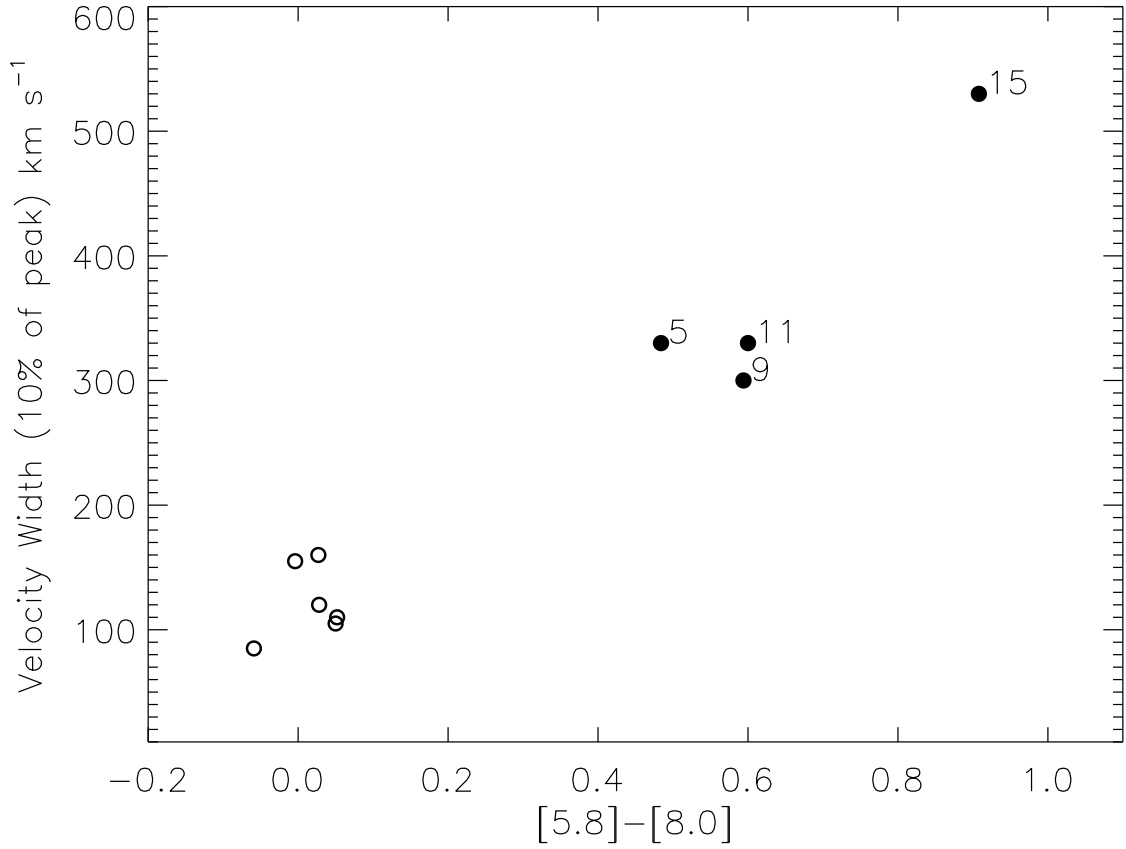}
\caption{Infrared excess as a function of accretion, using the
  velocity width of the H$\alpha$ line measured by \citet{law04} as a
  diagnostic of the accretion rate.  Stars with velocity widths
  greater than 270~km~s$^{-1}$ are accreting gas, the H$\alpha$
  emission in the remaining stars arises in active chromospheres
  \citep{white2003}. The filled circles are IRAC-excess sources, the
  open circles do not show excesses. }
\end{figure}

\clearpage

\end{document}